\documentclass[aps,prb,footinbib,superscriptaddress,twocolumn,showpacs,floatfix]{revtex4-1}

\usepackage{amsbsy}
\usepackage{amsmath}
\usepackage{amssymb}
\usepackage{amsfonts, dsfont}
\usepackage{pst-all}
\usepackage{epsfig}
\usepackage[english]{babel}
\usepackage{color}
\usepackage[normalem]{ulem}
\usepackage{bm}
\usepackage{yfonts}
\usepackage{comment}
\usepackage{balance}

\makeatletter
\newcommand*{\balancecolsandclearpage}{%
  \close@column@grid
  \clearpage
  \twocolumngrid
}
\makeatother

\begin{document}

\title{Energetics of Pfaffian-AntiPfaffian Domains}
\author{Steven H. Simon}
\affiliation{Rudolf Peierls Centre for Theoretical Physics, 1 Keble Road, Oxford, OX1 3NP, UK}

\author{Matteo Ippoliti}
\affiliation{Department of Physics,
Princeton University, Princeton, NJ 08544, USA}
\affiliation{Department of Physics, Stanford University, Stanford, CA 94305, USA}

\author{Michael P. Zaletel}
\affiliation{Department of Physics, University of California, Berkeley, CA 94720, USA}

\author{Edward H. Rezayi}
\affiliation{Department of Physics, California State University Los Angeles, Los Angeles, CA 90032, USA}

\date{\today}

\begin{abstract}
    In several recent works it has been proposed that, due to disorder, the experimentally observed $\nu=5/2$ quantum Hall state could be microscopically composed of domains of Pfaffian order along with domains of Anti-Pfaffian order.   We numerically examine the energetics required for forming such domains and conclude that for the parameters appropriate for recent experiments, such domains would not occur.  
\end{abstract}

\maketitle

Understanding the $\nu=5/2$ fractional quantum Hall states has been an enduring challenge for the condensed matter community.  Discovered experimentally\cite{Willett87} in 1987, the first compelling numerical work roughly a decade later\cite{Morf} identified this state as being the Moore-Read\cite{Moore}, or ``Pfaffian", phase of matter.    Almost another decade later it was realized that the particle-hole conjugate of the Pfaffian, the so-called ``AntiPfaffian", a different phase of matter, would fit the numerical data just as well\cite{LeeAntiPfaffian,Levin}.  Only very recently numerics have confirmed that the AntiPfaffian is likely to be energetically favorable over the Pfaffian for experiments in GaAs quantum wells\cite{RezayiRecent,RezayiSimon,Zaletel}. 

This numerical conclusion has been cast into doubt by recent thermal transport experiments\cite{Banerjee2018} which, at face value, do not match  predictions for either the AntiPfaffian or the Pfaffian.   This has raised the 
possibility\cite{Mross,Ashvin,Lian} that the observed quantum Hall state at filling fraction $\nu=5/2$ is microscopically composed of domains of Pfaffian order and domains of AntiPfaffian order, stabilized by disorder\cite{Ashvin}.   
Such mixed domains could have quantized electrical conductivity, identical to what one would observe for either the Pfaffian or AntiPfaffian phase, but could have either unquantized thermal conduction (a so-called ``thermal metal"\cite{ThermalMetal}) or, possibly with fine tuning\cite{Ashvin}, could result in a number of novel phases of matter including the PH-Pfaffian\cite{Son,Fidkowski}, the $K=8$ state, and the $113$ state\cite{Mross,Ashvin}.     
Although several alternative proposals\cite{SimonPaper,Feldman2,SimonRosenow} have been put forward to explain the experimental observations, the idea of multiple domains is certainly a compelling possibility. 

In the current paper we numerically examine the energetics of the Pfaffian and AntiPfaffian.    We derive the conditions that would result in domains in the physical system and compare them against what is known in the experiment.  Our results make the domain scenario appear impossible. 

The physical picture that could result in multiple domains is well described in Ref.~\onlinecite{Ashvin}.   It is known that at $\nu=5/2$, without Landau level mixing, the Pfaffian and AntiPfaffian are energetically equivalent.  
Landau level mixing weakly breaks this degeneracy and favors the AntiPfaffian\cite{RezayiRecent,RezayiSimon,Zaletel} for GaAs quantum wells similar to those of the experiments\cite{Banerjee2018}.   However, the quasielectrons of the Pfaffian and quasielectrons of the AntiPfaffian need not have the same energy.    
If the quasielectrons of the AntiPfaffian cost more energy than the quasielectrons of the Pfaffian, then when the filling fraction is increased enough the Pfaffian will be favored.
Conversely if the quasielectrons of the Pfaffian cost less energy than the quasielectrons of the AntiPfaffian, then due to the approximate particle-hole symmetry of the system, the quasiholes of the Pfaffian will have lower energy than the quasiholes of the AntiPfaffian, and as a result the Pfaffian will become favored when the filling fraction is lowered enough.     
Thus in either case, for a disordered system where the filling fraction is not uniform, one could obtain domains of Pfaffian and AntiPfaffian.  

Our first objective is to determine at what filling fraction the transition occurs from AntiPfaffian to Pfaffian.   To do this, we want to determine the energy of the quasiparticles of the Pfaffian versus the quasiparticles of the AntiPfaffian.   
Our result (see {\it Numerical-methods-1} section below) is that the quasielectrons of the Pfaffian have lower energy and the energy difference is approximately 
\begin{equation}
     E_{2qe} \approx  0.004  
     \label{eq:E2qp}
\end{equation}
where the energy here and elsewhere in the paper is given in natural units of $E_{interaction} = e^2/(\epsilon \ell)$ with $\ell$ the magnetic length.  This is the energy difference in removing one flux from the system for the two different wavefunctions.   
This result has been obtained using DMRG techniques  as explained in the methods section below.    In the methods section we also discuss the result of trying to estimate this quantity using variational and exact diagonalization techniques on smaller system sizes.

The fact that the quasielectrons of the Pfaffian are lower energy tells us that the putative transition from AntiPfaffian to Pfaffian occurs at $\nu > 5/2$.  
Thus for all $\nu<5/2$ we should have AntiPfaffian order with quasiholes.  This, in itself, is a significant result in the context of the experimental observation\cite{Kumar} of $\nu=2+6/13 < 5/2$ which can be easily interpreted as a daughter state of the AntiPfaffian via the mechanism of Ref.~\onlinecite{LevinHalperin} but not of the Pfaffian. 

We now want to compare this energy difference for quasielectrons to the energy difference between Pfaffian and AntiPfaffian at exactly $\nu=5/2$ due to Landau level mixing. 
Based on several recent works\cite{RezayiRecent,Zaletel} (See {\it Numerical-methods-2} section below) we establish an energy difference per electron between Pfaffian and AntiPfaffian of
\begin{equation}
  E_0 \approx .00066 \kappa
  \label{eq:E0}
\end{equation}
where this energy is per electron in the valence Landau level and $\kappa = E_{interaction}/(\hbar \omega_c)$ (with $\omega_c$ the cyclotron frequency) is the Landau level mixing parameter which is approximately 1.6 in the experiments of Ref.~\onlinecite{Banerjee2018}.

We now balance the energies of the Pfaffian with quasielectrons to that of the AntiPfaffian with quasielectrons to find the critical filling fraction where the transition occurs.   
Recall that $\nu = n_e \phi_0/B$ with $n_e$ the electron density and $\phi_0 = 2  \pi \hbar/e$, and we will consider only electrons in the valence Landau level (so we are taking $\nu=1/2 + \delta \nu$).    
The magnetic field is  $B = \nu^{-1} n_e \phi_0$, and the missing flux density compared to $\nu=1/2$ is
$
B_{1/2}-B = (2 -\nu^{-1}) n_e \phi_0
$. The energy difference per unit area between the quasielectrons of Pfaffian and AntiPfaffian is thus 
$
 E_{2qe} (2-\nu^{-1}) n_e 
$
whereas the energy difference per unit area from Landau level mixing is 
$
 E_0 n_e
$.
Thus the total energy difference goes through zero at the critical filling $\nu_c$ such that
$
 E_0 = E_{2qe} (2-\nu_c^{-1}),
$
or 
$$
 \nu^{-1}_c = 2 - E_0/E_{2qe}  \;.
$$
Plugging in numbers here gives us (at $\kappa=1.6$ appropriate for the experiments)
$$
 \nu_c \approx 0.58
$$
in the valence Landau level.  

This critical value can be seen to be problematic for the interpretation of the experiment\cite{Banerjee2018} as being made of domains of Pfaffian and AntiPfaffian.   In short, it requires an unreasonably large change in local filling fraction to putatively stabilize the Pfaffian.   
In the experiment, the quantum Hall plateau extends only between filling fractions  $\nu \approx 0.48-0.51$ within the valence Landau level.  Further, between $\nu=0.5$ and  $\nu_c=0.58$ there are several other phases of matter observed, including reentrant incompressible (presumed bubble) phases and compressible phases.   

Let us discuss the microscopic picture in a bit more detail.   
In a model with long range disorder, the system will break up into compressible and incompressible strips situated around the electrons' equi-density contours (which are in turn determined by the equi-potential contours of the disorder). 
Typically only a single such contour can percolate across the system, as regions of higher (lower) density are confined around local minima (maxima) of the disorder.
When this contour has filling fraction $\nu=0.48-0.51$ within the valence Landau level, the percolating state is incompressible, giving rise to a quantized Hall plateau.   
However, for filling fractions outside of the plateau, we should conclude that the phase at the percolating contour is compressible.   
Since $\nu_c$  is {\it far} outside of the plateau, we should assume that compressible regions (as well as possibly other states of matter) must intervene between AntiPfaffian order and any Pfaffian order, should the latter exist.   
At least in this picture of long range disorder (precisely the picture used to justify the domain scenario) it then seems impossible to construct an electrically incompressible phase of matter based on domains of Pfaffian and  AntiPfaffian.  Since our best estimate of $\nu_c$ is almost an order of magnitude further away from the center of the plateau than the half-width of the plateau itself, our conclusion should be very robust to any minor modification of assumptions (such as changes in methods of extrapolating to large system size, {\it etc};  see numerical methods sections). 

For a hypothetical system where Landau level mixing  (and therefore $E_0$) is much smaller, one could imagine a situation where the energetics of mixed domains is more viable.  If there were a case where Pfaffian domains are possible, a more accurate accounting of the full energy budget  would also include the energy costs associated with the tension of the domain wall itself.  This would set a minimum size of Pfaffian domains within the AntiPfaffian background.   Using DMRG we have estimated the domain wall tension to be on the order of $0.0010 - 0.0024 \, e^2/{\epsilon \ell^2}$.  In the case of the experiments of Ref.~\onlinecite{Banerjee2018}, this tension, not previously considered, makes the idea of mixed domains even less favorable.  The detailed energetics of domain walls are discussed in the Supplmemental material to this paper.\cite{Supplement}

To summarize the result of our paper:  We find that for typical high mobility samples similar to that of Ref.~\onlinecite{Banerjee2018}, the picture of mixed domains of Pfaffian and AntiPfaffian proposed by Refs.~\onlinecite{Mross,Ashvin,Lian}  is not viable.

\begin{figure}
    \centering
    \includegraphics[width=\columnwidth]{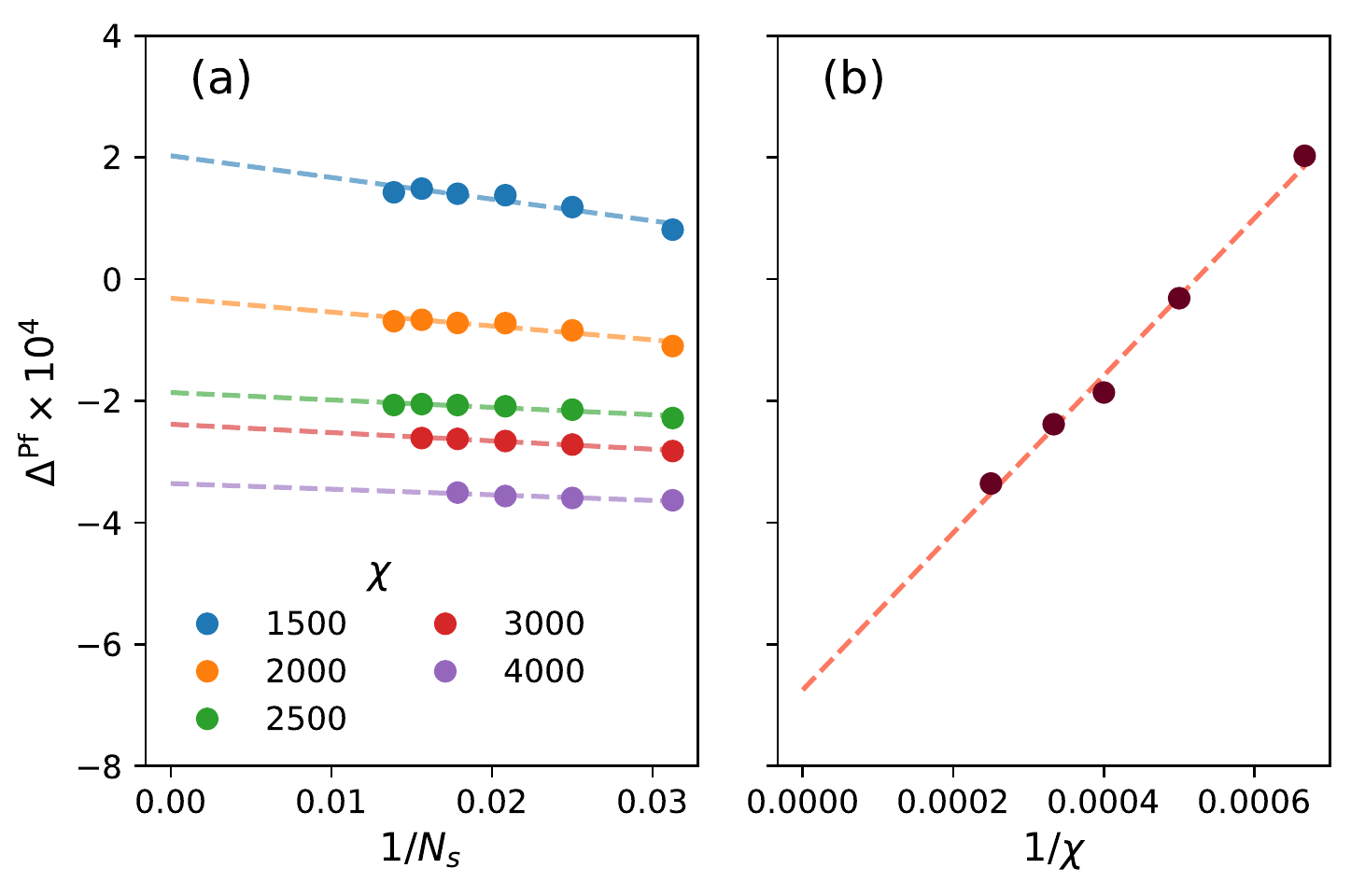}
    \caption{Example of extrapolation of DMRG results in the two cutoffs $N_s$ and $\chi$. 
    (a) Extrapolation in the size of the interface region $N_s$. 
    (b) The $N_s\to\infty$ values are extrapolated in the bond dimension $\chi$.
    The data is for the Pfaffian wavefunction at $L=26\ell$.}
    \label{fig:extrap}
\end{figure}

{\it Numerical Methods 1:  Quasielectron vs Quasihole Energy.}    Our numerical method is based on DMRG on an infinite cylinder geometry.
We first generate Pfaffian and AntiPfaffian matrix product state (MPS) wavefunctions by using DMRG on the half-filled $N=1$ LL, for several values of the cylinder circumference $L$ and MPS bond dimension $\chi$ (larger $\chi$ being more accurate).
By seeding the DMRG with product states with different ``root configurations,'' we are able to obtain all the distinct topologically degenerate ground states  of both the Pf and APf phases.
To determine the quasielectron, quasihole, and domain wall energies, we use these different ground states as semi-infinite left / right boundary conditions in a \emph{finite} DMRG variational optimization on a number $N_s$ of orbitals, as described in Ref.~\onlinecite{ZaletelMongPollmannPRL}.
By choosing different topological sectors $a, b$ for the left / right boundary condition, we form an interface between two distinct topological sectors, which traps an anyon $c$ in the fusion outcome $N^c_{a \bar{b}}$.
For example, using sector $a = \sigma^+$ (root $\cdots 0101 \cdots$, where 0 and 1 denote empty and occupied Landau orbitals in the initial product state) on the left and  sector $b = \mathds{1}$ (root $\cdots 0110\cdots$) on the right gives the $+e/4$ quasielectron $\sigma^+$. Exchanging left and right gives the $-e/4$ quasihole.
From this we obtain the energies $E_{\pm e/4}^\alpha$, with $\alpha = \text{Pf}$ or $\text{APf}$, relative to that of the vacuum.
We calculate the energies for multiple values of the cutoffs $32 \leq N_s \leq 80$ and $1500 \leq \chi \leq 4000$ and extrapolate both to infinity, as shown in Fig.~\ref{fig:extrap}.

\begin{figure}
    \centering
    \includegraphics[width=\columnwidth]{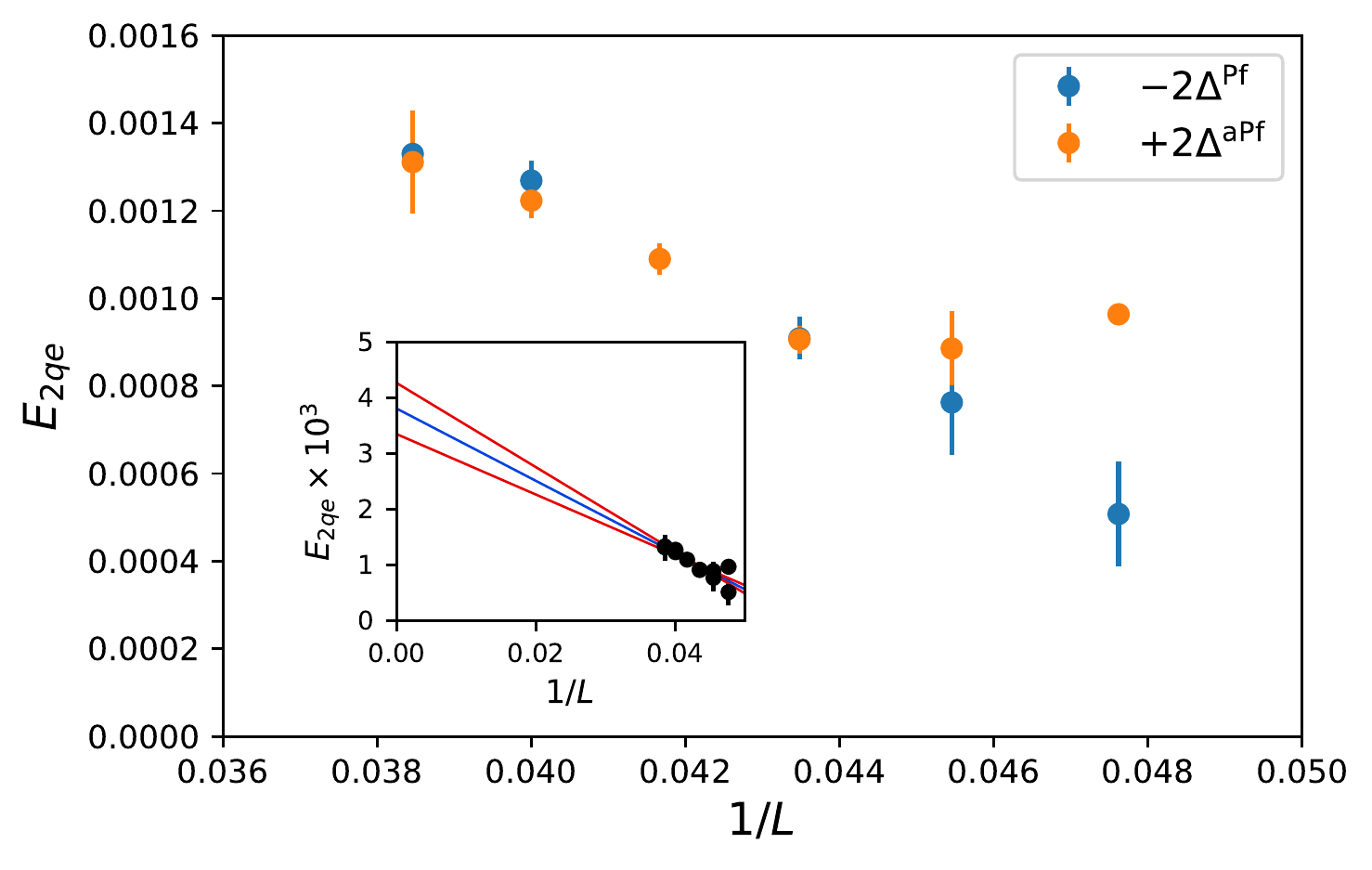}
    \caption{DMRG results after extrapolation in the cutoffs. 
    $
    \Delta \equiv E_{+e/4} - E_{-e/4} - \frac{1}{2} E(1)
    $, 
    for both Pfaffian and AntiPfaffian, plotted vs the inverse cylinder circumference $1/L$ (in units of $\ell$). 
    We find $\Delta^\text{Pf}<0$ (i.e. the Pfaffian has lower-energy quasielectrons) and the PH relation $\Delta^{\rm Pf}+\Delta^{\rm APf}=0$ is obeyed quite accurately for $L\geq 23\ell$.  Inset: $1/L$ extrapolation of the data (outer lines indicate the uncertainty of the fit).}
    \label{fig:delta}
\end{figure}

The quantity of interest is 
$
E_{2qe} = 2(E^\text{APf}_{+e/4} - E^\text{Pf}_{+e/4})
$, 
but it is convenient to define 
$
\Delta^\alpha \equiv E^\alpha_{+e/4} - E^\alpha_{-e/4} - \frac{1}{2} E(1)
$, 
where $E(1)$ is the energy of a filled $N=1$ LL. 
These satisfy 
$
\Delta^\text{APf} - \Delta^\text{Pf} = E_{2qe}
$ 
and 
$
\Delta^\text{Pf} + \Delta^\text{APf} = 0
$ 
due to particle-hole symmetry; the latter provides a numerical consistency check.
The extrapolated values of $\Delta^\alpha$ are shown as a function of cylinder circumference 
$
21 \ell \leq L \leq 26\ell
$ 
in Fig.~\ref{fig:delta}.
At the available sizes we find 
$
E_{2qe} \simeq 0.001
$.
While the value appears to drift with $L$, a linear extrapolation in $1/L$ (likely an overestimate) yields 
$
E_{2qe} \simeq 0.004
$, 
which is the result we use in Eq.~\ref{eq:E2qp}.  
Note that other reasonable extrapolations could easily have given a much lower value (whereas a much larger value appears very unlikely).   
A smaller estimate of $E_{2qe}$ would only give a {\it larger} value of $\nu_c$ making our conclusion even stronger.
The Pfaffian-AntiPfaffian domain wall tension can similarly be computed by forming an interface between their respective vacuum sectors. 

\begin{figure}[t]
    \centering
    \includegraphics[width=0.8 \columnwidth]{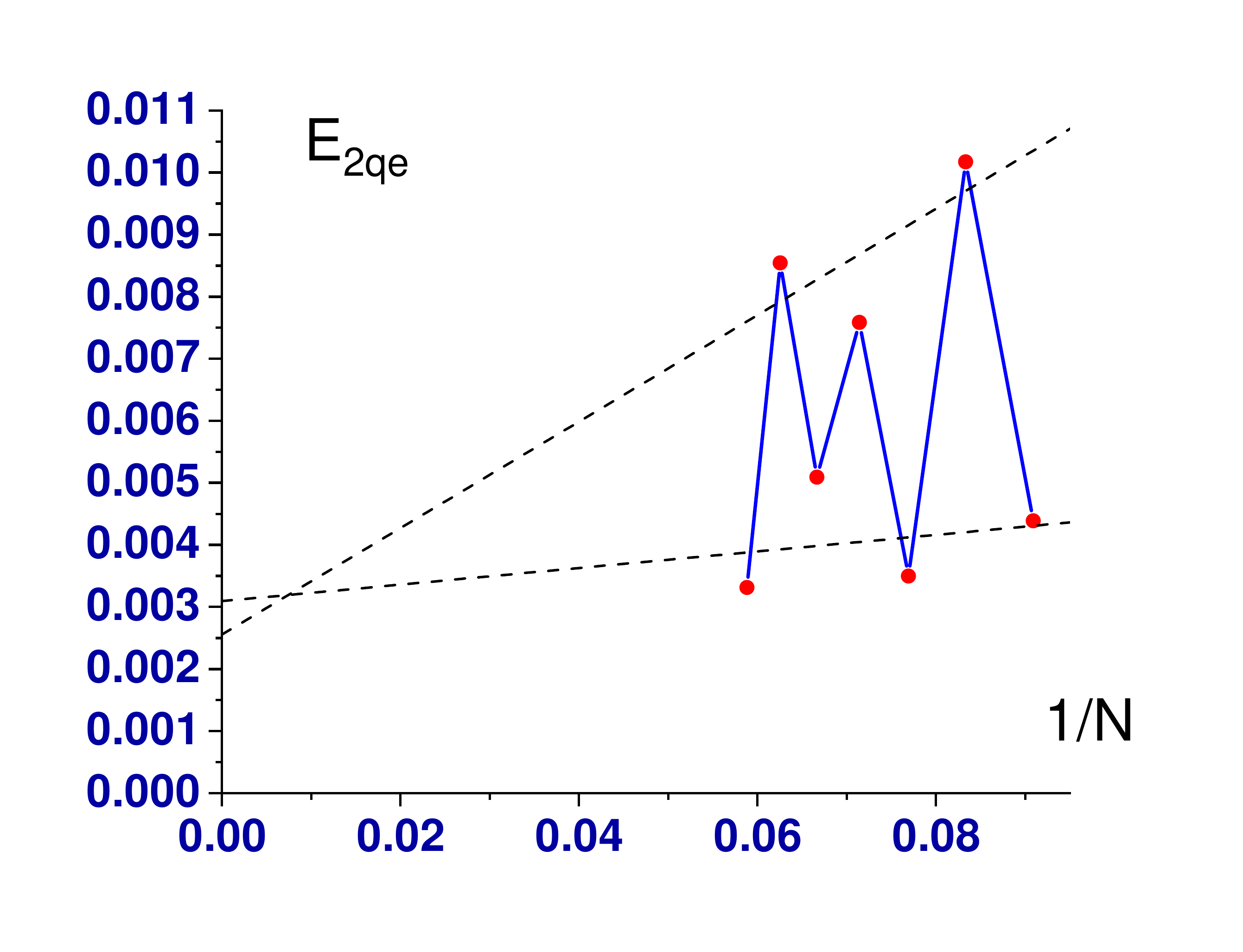}
    \caption{Quasielectron energy difference between Pfaffian and AntiPfaffian as a function of system size.  The numerical method is variational and exact diagonalization (See text). The upper points are for even numbers of electrons and the lower points are for odd.    The fact that the two linear extrapolations do not intersect at $1/N=0$ suggests that we cannot completely trust the extrapolation.
    Here, $N=11,12,13,14,15,16,17$ on a hexagonal torus.
}
    \label{fig:Edpic}
\end{figure}

To support our results we also use the exact Pfaffian and AntiPfaffian wavefunctions on finite systems to obtain variational estimates of quasielectron and quasihole energies at a total flux that is displaced by $\pm \phi_0$ away from $\nu=5/2$. 
We generate the Pfaffian by an ultra-short-range 3-body repulsive potential $H_\text{Pf}$. 
For our purposes it is more convenient to generate the AntiPfaffian by a potential that is the particle-hole conjugate of the $H_\text{Pf}$, rather than taking particle-hole conjugates of a series of wave functions. 
For odd number $N$ of electrons, we find an equal number $(N + 1)/2$ of degenerate states for quasielectrons and quasiholes.
Particle-hole conjugation in finite-sized systems at these fillings changes the electron number by 1. 
As a result, for even electrons, the number of degenerate quasiholes of the Pfaffian exceed those of the AntiPfaffian quasielectrons by 1.
 All the sizes presented in Fig.~\ref{fig:Edpic} 
involve quasielectrons and are unrelated by particle-hole conjugation. 
The quasielectrons of the Pfaffian (as well as the quasiholes of the Antipfaffian) are generated by their respective Hamiltonians and are not degenerate at these fillings.
In these cases we find the lowest variational energy of the second Landau level Coulomb interactions among the lowest $(N + 1)/2$ model states (typically the first, occasionally the second if the lowest two are close in energy). 
In cases where we have degeneracies we find the optimum state of quasielectrons  by diagonalizing the Coulomb potential in the subspace of the degenarate manifold of model states. 
We take the lowest energy states of the multiplets (corresponding to the quasielectrons maximally spaced from each other). 
Since we are limited to smaller sizes it is harder to draw definitive conclusions.
 It is clear from the data that the extrapolation is
likely to have a substantial uncertainty. However, we use
these as guides to estimate bounds on the energy. 
From the data shown in Fig.~\ref{fig:Edpic} the extrapolation of the value
of $E_{2qe}$ is somewhere between 0.0025 
and 0.0062, which is highly consistent with our DMRG value. 
The lower bound makes $\nu_c$ larger than the DMRG's. 
Even if we take the higher number seriously, this only reduces $\nu_c$ to .54, which is still far outside the experimentally observed plateau.

To avoid getting too close to phase 
boundaries across which a transition to a crystalline phase may occur, 
we have altered the Coulomb interaction  by a small amount. 
For all sizes above $N=12$, as in our DMRG calculation,  we have added 0.0325 to the
$v_1$ pseudopotential.  
For our largest size $N=17$ system that we have studied by exact calculations, the first two pseudopotentials of the Coulomb 
repulsion for electrons 
in the second Landau Level are $v_1=0.4246310$ and $v_3=0.3306233$ (only odd relative angular momenta  are relevant since the valence electrons are believed to be fully spin-polarized).
 For smaller sizes we have instead made a smaller change of 0.0225. This compensates 
for the fact that smaller sizes have larger $v_1$ and thus avoids a 
transition to the composite Fermi liquid phase.  

{\it Numerical Methods 2: Pfaffian-AntiPfaffian Splitting.}  We start by examining the iDMRG results  from  Ref.~\onlinecite{Zaletel}.   
The first key result (Fig.~7 of that work) is that the energy splitting between Pf and APf is very nearly linear in Landau level mixing up to at least $\kappa=1.38$.   The magnitude of the splitting in that plot is roughly .00047 $\kappa$.  However, upon increasing the number of Landau levels included in the calculation, the spitting increases (Fig.~9 of that work).   The authors of that work warn us not to take this result to be too precise quantitatively, but nonetheless we are able to trust the trends.  We then examine the diagonalization results of Ref.~\onlinecite{RezayiRecent}.  Here the splitting is calculated to first order in $\kappa$.   However, given the linearity with $\kappa$ obtained in Ref.~\onlinecite{Zaletel}, this appears to be sufficient.  From the the inset to Fig.~1 in Ref.~\onlinecite{RezayiRecent}, extrapolated to large system size, we obtain the stated energy splitting $.00066 \kappa$ that we use in Eq.~\ref{eq:E0}.    Note that there may be some uncertainty in this number given that we are extrapolating to large $\kappa$ and large system size.   However, as mentioned in the main text, even a very substantial change in this number would not change the final conclusion of our paper. 

Acknowledgments:  SHS is supported by EPSRC grants EP/S020527/1	and  EP/N01930X/1.  Statement of compliance with EPSRC policy framework on research data:  This publication is theoretical work that does not require
supporting research data. ER and MI were supported by DOE grant No.~DE-SC0002140. MI was funded in part by the Gordon and Betty Moore Foundation's EPiQS Initiative through Grant GBMF4302 and GBMF8686. MZ is supported by ARO MURI grant No.~W911NF-17-1-0323.

\bibliographystyle{apsrev4-1}
\balance
\bibliography{test}

\balancecolsandclearpage

\section*{Supplemental Material:  Domain Wall Tension Energetics}


In this supplement we imagine a system where the energetics are more favorable for domain formation (perhaps due to lower Landau level mixing) and we consider the energy of a domain wall.   
Here we assume that due to local density disorder the system will be partitioned into regions of Pfaffian and regions of AntiPfaffian. 

Let us consider a circular domain of radius $R$ which we measure in units of the magnetic length.  Let us assume  that the filling fraction outside of the domain is below $\nu_c$ so we have AntiPfaffian, but inside the domain wall the filling fraction is greater than $\nu_c$ so that if the domain is large enough, the phase inside will be Pfaffian.

As mentioned in the main text, the energy difference per electron between Pfaffian and AntiPfaffian inside the putative domain is 
$
E_{2qe} (2-\nu^{-1})  - E_0 
$.
Having set the magnetic length to unity we have $\nu = n_e 2 \pi $.
So in a circular domain of radius $R$ we have a total of
$
 N = (\pi R^2) \frac{\nu}{2 \pi} 
$
electrons.  
The energy gain of the domain is thus
$$
\left( E_{2qe} (2-\nu^{-1})  - E_0  \right) \nu R^2/2 \;,
$$
which we balance against the domain wall energy.   Let $\tau$ be the domain wall tension,
i.e. its energy per unit length (in units of $e^2/\epsilon \ell^2$).  
As mentioned in the main text, a DMRG calculation gives this domain wall tension 
$\tau =$ 0.001 -- 0.0024. 
The circumference is $2 \pi R$ so the domain wall energy is
$
 \tau 2 \pi R
$.
Thus we have 
$$
 R =  \frac{4 \pi \tau} { E_{2qe} (2\nu-1)  - \nu E_0 }
$$
as the minimum domain radius.

Finally we have to input what the filling fraction is inside our domain.    Obviously if $\nu$ is only a very small amount above $\nu_c$, then Pfaffian and AntiPfaffian will be very close to degenerate, and the minimal domain size will need to be very large to overcome the domain wall energy (diverging as $R\sim (\nu-\nu_c)^{-1}$).   
However, if $\nu$ is much bigger than $\nu_c$ then there is a large energy gain to having a region of Pfaffian order and even a small domain can overcome the wall tension.

 One should realize that this crude estimate assumes $\nu$ is constant everywhere inside the domain, and this is certainly not going to be the case -- accounting for this correction will make the minimal radius of the domain even larger.

{\it Numerical Methods For Domain Wall Energy.}   Here we again use real-space DMRG on a cylinder with the boundary condition  set to be the Pfaffian wavefunction  on one side and the AntiPfaffian wavefunction on the other (both in the $\mathds{1}$ sector), and perform a variational optimization on $N_s$ orbitals in the middle.  
Landau level mixing is turned off here. 
The ground state energy of the system with these boundary conditions is compared against the energies of the infinite Pfaffian (or AntiPfaffian) wavefunctions; the energy difference is the cost of the domain wall.
Again we must extrapolate two cutoffs (the lenght of the segment $N_s$ and the bond dimension $\chi$) to infinity, as in Fig.~\ref{fig:extrap}.
We extract the domain wall tension from a linear fit in the cylinder circumference $L$, finding values between $0.001$ and $0.0024$ depending on the form of the interaction.

\end{document}